\documentclass[11pt]{article}

\usepackage{graphicx}
\usepackage{amsmath,amssymb,amsthm}
\usepackage{lscape}
\usepackage{placeins}
\usepackage{hyperref}
\usepackage{verbatim}
\usepackage[margin=1.5in]{geometry}
\usepackage[]{algorithm2e}

\newtheorem{definition}{Definition}
\newtheorem{condition}{Condition}

\usepackage{xcolor}
\usepackage{soul}

\title{The Opacity Problem in Social Contagion}

\author
{George Berry$^{a\ast}$, Christopher J. Cameron$^{a}$,\\Patrick Park$^{b}$, and Michael Macy$^{a,c}$\\
\\
\normalsize{$^{a}$Department of Sociology, Cornell University,}\\
\normalsize{323 Uris Hall, Ithaca NY 14853, USA}\\
\\
\normalsize{$^{b}$SONIC Research Group, Northwestern University,}\\
\normalsize{Frances Searle Building, Evanston, IL 60208, USA}\\
\\
\normalsize{$^{c}$Department of Information Science, Cornell University}\\
\normalsize{Gates Hall, Ithaca NY 14853, USA}\\
\\
\normalsize{$^\ast$To whom correspondence should be addressed; E-mail: \href{mailto:geb97@cornell.edu}{geb97@cornell.edu}.}
}

\begin{document}

\maketitle

\begin{abstract}

Many social phenomena can be modeled as cascades in networks, where nodes adopt a behavior in response to peers adopting. When studying cascades, researchers typically assume that the number of active peers when a node adopts is equivalent to the node's threshold for adoption. This assumption is rarely justified due to the ``opacity problem'': networked cascades reveal intervals which contain thresholds, rather than point estimates. Existing approaches take the maximum of each node's threshold interval, which biases models of social influence. Opacity is inevitable in many small graphs when using the threshold model, resulting from the networked process itself rather than data collection techniques. Using simulation, we extend this finding to the probabilistic SI (independent cascade) model. We confirm these theoretical results by studying 50 large hashtag cascades among 3.2 million Twitter users, finding that 20\% of adoptions suffer from opacity. Different assumptions in response to opacity qualitatively change conclusions about peer influence. While opacity is a far-reaching problem, it can be addressed. Using information from nodes who have tightly bounded intervals allows building models to reduce error in estimating node thresholds.
\end{abstract}


\break

\section{Introduction}


Like epidemic diseases \cite{klovdahl_social_1994,van_den_hof_measles_2001,campbell_complex_2013}, social contagions are ubiquitous, highly consequential, and widely studied \cite{granovetter_threshold_1978,strang_diffusion_1998,strogatz_exploring_2001,centola_complex_2007,macy_chains_1991,valente_network_1995,valente_social_1996,christakis_spread_2007,valente_network_2012}. Unlike nearly all diseases, social contagions often require multiple sources of infection, especially if adoption is costly, risky, motivated by affect, or entails positive network externalities \cite{centola_complex_2007}.

In this paper, we explore a challenge for studies of contagion: data generated by networked processes typically does not provide all of the information needed to recover peer influence, even when cascades are tracked precisely. This occurs due to the ``opacity problem'': cascades provide intervals in which node thresholds lie, but often do not indicate the threshold itself. Researchers typically infer the amount of social reinforcement required for adoption by recording node exposure (number of active neighbors) when the node activates. We refer to this practice as the exposure-at-activation (EAA) rule, which is equivalent to taking the maximum of the threshold interval for each node. This leads to an upward bias in estimates of peer reinforcement needed for adoption at the node level, and can bias models of peer effects.

The EAA rule can be found in diverse studies across disciplines, including sociology \cite{friedkin_social_1990,valente_social_1996,dimaggio_network_2012}, economics \cite{de_giorgi_identification_2010,bramoulle_identification_2009,banerjee_diffusion_2013}, medicine \cite{christakis_spread_2007}, and information science \cite{aral_distinguishing_2009,romero_differences_2011,steeg_what_2011,gonzalez-bailon_dynamics_2011,ugander_structural_2012,borge-holthoefer_cascading_2013,weng_virality_2013}. It is even implicitly present in ``snapshot'' studies \cite{backstrom_group_2006, crandall_feedback_2008} and in networks surveyed in waves such as Add Health \cite{bearman_suicide_2004}.

Simulated cascades using both deterministic and probabilistic activation rules indicate that the opacity problem creates substantial bias. An empirical study of hashtag cascades among 3.2 million users on Twitter confirms this intuition: about 20\% of hashtag first usages have uncertain adoption thresholds. For 2 in 5 hashtags, different responses to the opacity problem produce qualitatively different conclusions about whether peer reinforcement promotes diffusion.

We provide several tools for addressing the opacity problem. First, a simple condition can be applied to recover node threshold intervals. Nodes with small threshold intervals can be used to estimate a threshold model, which can then be applied to all nodes. In simulations, this process substantially reduces error in estimating thresholds. In addition, we expect future research to develop novel ways to further reduce measurement error introduced by opacity and the EAA rule.

While the opacity problem may seem complex, the core intuition is simple and is summarized in Figure \ref{fig111}: for some network-threshold configurations, some nodes will always activate with exposure greater than threshold. This occurs because all nodes cannot check the status of neighbors all of the time. To use a familiar example: suppose a person ``looks away'' from her phone while many friends adopt a behavior on a social media platform. When she comes back, she sees that several friends have adopted and she promptly adopts. Ascertaining which friend was pivotal is a complex counterfactual question which the data does not directly answer. A similar analogy can be made about states adopting policies \cite{walker_diffusion_1969}, college students selecting majors \cite{de_giorgi_identification_2010}, or any number of other behaviors.

The paper proceeds as follows: In Section 2, we introduce the threshold model and show that for some specific network configurations, the opacity problem is inevitable. An exhaustive survey of all small graphs with all threshold assignments is conducted, which indicates that even in small networks, the opacity problem creates substantial uncertainty. We propose a condition for bounding node-level uncertainty which is the basis of many of our subsequent results. In Section 3, we conduct a variety of simulations using the threshold model from Section 2, finding that the application of the EAA rule creates substantial upward bias in estimates of thresholds. We also simulate the susceptible-infected (SI) \cite{hethcote_mathematics_2000} or independent cascade model (ICM) \cite{kempe_maximizing_2003}, finding that the opacity problem arises in the probabilistic case as well. We show that a modeling procedure using node threshold intervals with low uncertainty can substantially reduce the upward bias of the EAA rule. In Section 4 we turn to an empirical case: hashtag cascades on Twitter. This analysis indicates that the opacity problem is present in a data-rich case from social media. We conclude with a discussion of implications and suggestion for future work.

To our knowledge, we are the first to highlight the full scope of this problem, although aspects of it have been discussed in \cite{valente_social_1996,leskovec_dynamics_2008,romero_differences_2011}.

\FloatBarrier

\section{Model}

\subsection{Model definition}

We choose a threshold model similar to previous work \cite{granovetter_threshold_1978,watts_simple_2002, kempe_maximizing_2003} to illustrate the opacity problem. We study the threshold model formally, and show using simulation below that our results apply to probabilistic models of contagion as well.

A graph $G = (V, E)$ has nodes $V$ and edges $E$. We assume that $G$ is undirected and connected. Nodes are indexed by $i$ and $j$, with $N(i)$ indicating all nodes in the neighborhood of $i$. A diffusion process plays out on this graph over times $t \ge 0$. Nodes have activation statuses at each $t$ indicated by $y_i(t) \in \{0, 1\}$, with $y_i(t)=0$ indicating inactive and $y_i(t)=1$ indicating active. Once a node activates, it remains active for the rest of the diffusion process. Edges may have weights $w_{ij}$, although weights are set to 1 in results presented here to reduce the complexity of the model space\footnote{Constructing a weighted example where opacity happens is straightforward. For example, change one of the edge weights to 2 in Figure \ref{fig111}.}.  Nodes have integer-valued thresholds $h_i \ge 0$, indicating the minimum level of peer reinforcement required for adoption\footnote{Results do not change with fractional thresholds, as demonstrated in simulations below.}. A node adopts when it updates at time $t$ and exposure is greater than threshold, $k_i(t) = \sum_{j \in N(i)} w_{ij} y_j^t \ge h_i$, where $j$ ranges over neighbors of $i$.

The update step proceeds as follows: at time $t$, an inactive node $i$ is chosen at random to update\footnote{Any continuous distribution producing node update times will give a full ordering of nodes with probability 1, since the probability of two nodes updating at exactly the same time is 0. We thank an anonymous reviewer for pointing this out. Simultaneous updating processes do not fix the opacity problem, as indicated in a previous draft of this paper.}. This update process produces a well-defined update ordering contained in vector $\mathbf{u}$. The vector $\mathbf{u}_i$ indicates all times at which $i$ updates. When $i$ updates, $i$ checks the status of all neighbors $j \in N(i)$, and if $\sum_{j \in N(i)} y_j \ge h_i$, $i$ activates immediately and $y_i(t) = 1$. While random updating and instantaneous activation are not realistic modeling choices, results are not sensitive to these simplifications\footnote{Figure \ref{fig111} shows a case which demonstrates this: the opacity problem occurs regardless of the order in which nodes update, meaning that random updating is not a critical assumption. Giving nodes an activation delay after updating would not substantively change results, since $j$ or $k$ would have to activate first, leading to the other node being measured with error.}.

In an important distinction from past work, whether or not $i$ activates at $t$, $i$'s exposure $k_i(t)$ is recorded. Recording $i$'s exposure when $i$ \textit{does not} update is crucial to addressing the opacity problem. Throughout the paper, we will refer to exposure-at-activation (EAA) and the ``EAA rule'', which we define here.

\begin{definition}
Assume node $i$ first activates at time $t$. Then $i$'s \textbf{exposure-at-activation (EAA)} is $k_i(t)$.
\end{definition}

\begin{definition}
The \textbf{exposure-at-activation (EAA) rule} estimates threshold $h_i$ by the exposure-at-activation. If $i$ first activates at time $t$, the EAA is $k_i(t)$ and the EAA rule assumes $h_i = k_i(t)$.
\end{definition}

The use of the EAA rule is straightforward when adoptions are timestamped. However, many studies of diffusion use networks surveyed in waves, such as Add Health (e.g. \cite{bearman_suicide_2004}) or network snapshots \cite{backstrom_group_2006}. The activation decision is only observed after a (potentially unknown) delay. In this case, the time a survey is administered is functionally the ``adoption time'', since it is the first time a node is observed as active. If ``survey time'' or ``collection time'' is substituted for ``activation time'', our results fully apply to network snapshots as well.

\subsubsection{Model remarks}

We have chosen a simple threshold model which abstracts away many aspects of reality. While past work has made similar simplifying assumptions \cite{granovetter_threshold_1978,kempe_maximizing_2003,kempe_influential_2005,centola_complex_2007,watts_simple_2002}, this paper fundamentally concerns empirical data. We use a model to motivate a claim about the empirical study of social contagion.

Because of this, it is crucial that results presented here are not artifacts of particular modeling assumptions. We have attempted to examine our results under as many assumptions as possible. Cases we have considered are: integer and fractional thresholds, weighted networks, non-random update orderings, various distributions of thresholds (normal, uniform, exponential), allowing nodes to update simultaneously, probabilistic activation models (independent probabilities), allowing nodes to have activation delays, allowing nodes to have transmission delays\footnote{We thank an anonymous reviewer for this suggestion.}, and allowing nodes to notify neighbors of activation. In each of these cases, the opacity problem remains a problem, and can be demonstrated with simple examples akin to Figure \ref{fig111}.

On the other hand, the threshold model considered here has a surprising amount of flexibility. It makes few assumptions about $G$ or the distributions of $h_i$ and $\mathbf{u}$. This is by design: the opacity problem occurs in small graphs that are very likely to occur for a wide range of specific assumptions.

While we consider several extensions to this model, four assumptions are fixed throughout. The most important of these is the static nature of the graph $G$. In reality, edges are created and decay in networks all the time \cite{kossinets_origins_2009}. While a dynamic graph does not eliminate the opacity problem, it may reduce its impact by sparsifying the graph. A second simplification is the assumption that diffusion is entirely endogenous: once the threshold is known, no other node-level information is relevant for the unfolding of the diffusion process. In practice, changing circumstances may dynamically alter node thresholds, creating additional challenges for estimating peer effects. Third, we assume throughout that peer influence is driving adoption, rather than a process orthogonal to the social network. Fourth, we assume that people are aware of the activation statuses of their neighbors and do not suffer from limited attention \cite{weng_competition_2012}\footnote{We thank an anonymous reviewer for suggesting this point.}.

\subsection{The opacity problem in the threshold model}

\begin{figure}[ht]
\includegraphics[width=\linewidth]{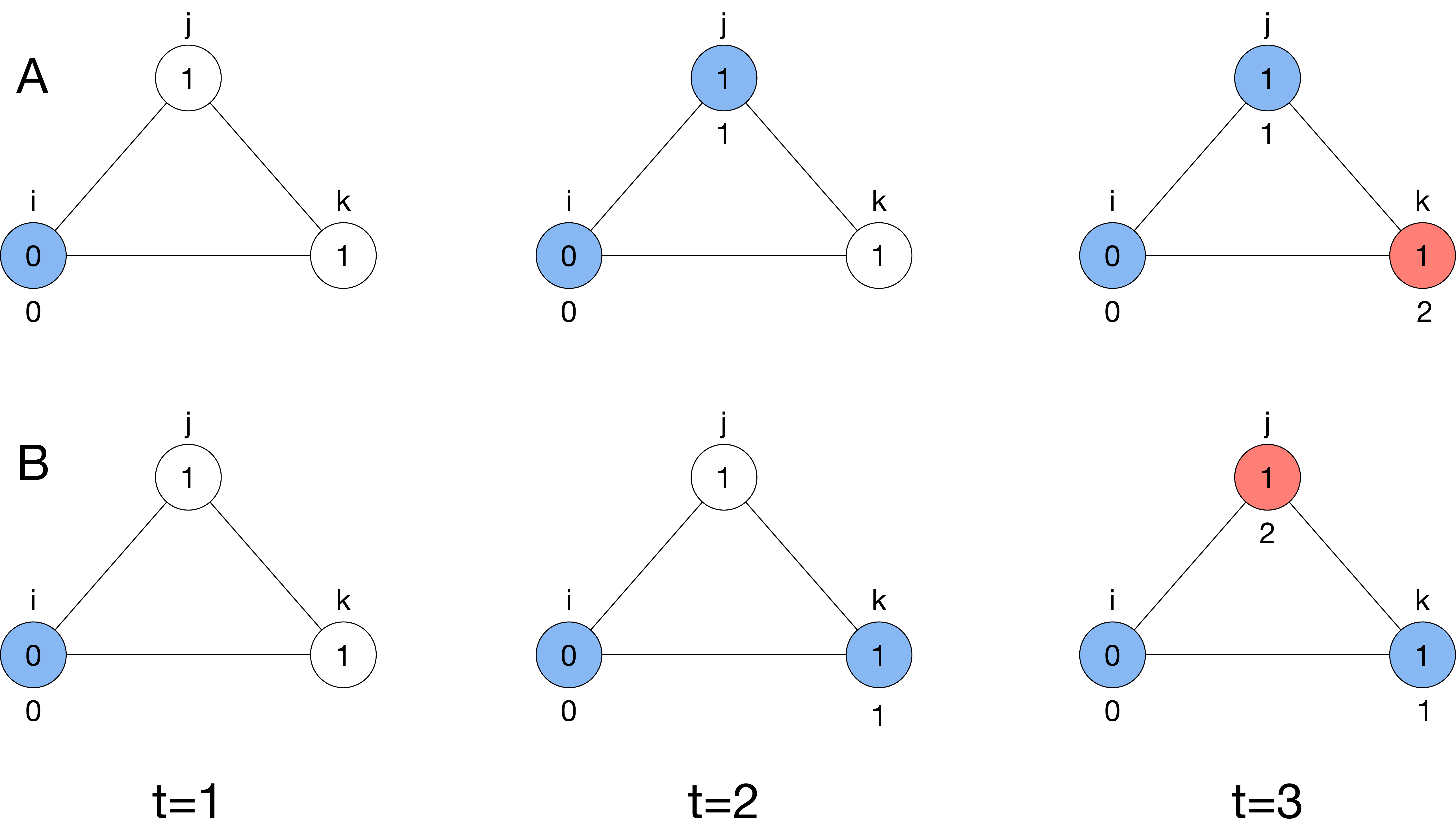}
\centering
\caption{A triad between nodes $(i, j, k)$ with thresholds $(0, 1, 1)$ over three time periods. A) and B) display the only possible activation orderings: $(i, j, k)$ or $(i, k, j)$, respectively. Node thresholds are displayed on the node, while the exposure at activation (EAA) is displayed below the node. Blue indicates a node activates with EAA equal to threshold, while red indicates that EAA is greater than threshold. Since these are the only two possible activation orderings, either $j$ or $k$ will always have EAA greater than threshold, leading to measurement error.}
\label{fig111}
\end{figure}

The opacity problem is inevitable in certain network configurations using the threshold model defined above. Consider the simple network in Figure \ref{fig111}, which is a triad among nodes $i, j, k$. Node $i$ is an innovator with threshold 0, while nodes $j$ and $k$ have threshold 1. Any update ordering for this graph will produce an over-estimate of at least one node's threshold. Since node $i$ must adopt first, there are only two possible orderings in which nodes can activate: $(i, j, k)$ or $(i, k, j)$. In the former case, $k$ adopts with 2 active neighbors despite having threshold 1, while in the latter case $j$ adopts with 2 active neighbors despite having threshold 1. Applying the exposure-at-activation (EAA) rule would produce an over-estimate of the node threshold in these cases.

In practice, how can we determine if node thresholds are measured precisely or not? For instance, in Figure \ref{fig111}-A, node $j$ could activate with 1 active neighbor but have a true threshold of zero. We need to place a lower bound on the threshold in addition to an upper bound. Recording exposure $k_i(t)$ when nodes remain inactive allows us to do this.

\begin{figure}[ht]
\includegraphics[width=\linewidth]{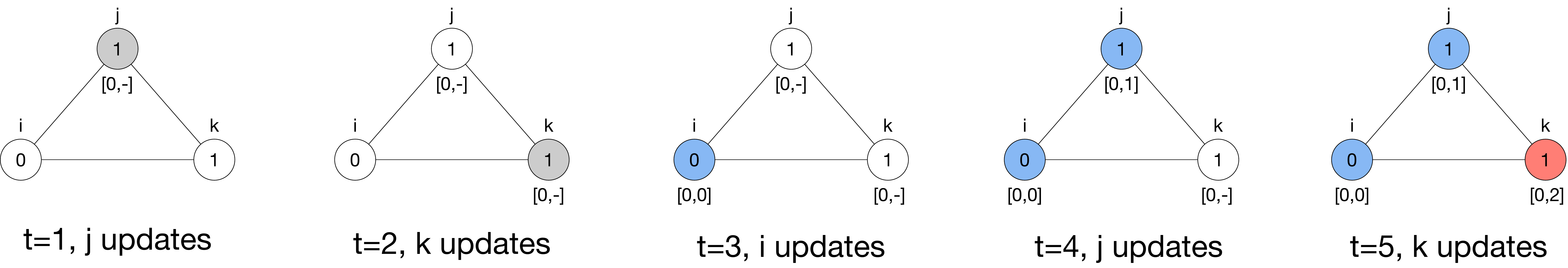}
\centering
\caption{A triad between nodes $(i, j, k)$ with thresholds $(0, 1, 1)$ over five time periods. Nodes update in order $(j, k, i, j, k)$. This ordering allows placing bounds on thresholds for each node, indicated by the intervals below each node (where ``-'' indicates no measurement). Gray indicates a node updates but does not activate, blue indicates a node activates with exposure-at-activation (EAA) equal to threshold, while red indicates a node activates with EAA greater than threshold. When nodes $j$ and $k$ update at $t = 1, 2$, respectively, their exposure is recorded even though neither activates. This allows placing a lower bound on their thresholds. As the process plays out, $k$ activates with EAA greater than threshold. The lower bound of 0 from $k$'s update at $t=2$ means $k$ has a threshold interval of $[0, 2]$.}
\label{fig222}
\end{figure}

Consider the diffusion process in Figure \ref{fig222}, which replicates the graph and threshold assignment of Figure \ref{fig111} but changes the update ordering to $\mathbf{u} = (j, k, i, j, k)$. This update ordering does not affect the order in which nodes activate (which remains $(i, j, k)$), but provides lower bounds on the threhsolds of nodes $j$ and $k$. The lower bound on node $i$ arises by assuming there are no negative thresholds\footnote{The appropriateness of this assumption has to be investigated in each individual case. Negative thresholds may indicate ``super-eager'' nodes.}. Note that node $j$ has an interval of $[0, 1]$, indicating that 0 active neighbors were insufficient to trigger adoption, while 1 active neighbor was sufficient. $j$'s threshold is therefore known with certainty to be 1 (recall we assume integer threhsolds). For node $k$, 0 neighbors was insufficient while 2 neighbors were sufficient, indicating uncertainty about whether $k$'s threshold is 1 or 2. Therefore, $j$'s threshold is measured precisely while $k$'s threshold suffers from opacity.

This suggests a formal precise measurement condition. Consider a node $i$ which updates at $t$ and some later $t'$. Assume $i$ is inactive at $t$ ($y_i(t) = 0$) and active at $t'$ ($y_i(t') = 1$), giving threshold interval $[k_i(t), k_i(t')]$.

\begin{condition}
The threshold for $i$ is \textbf{precisely measured} if $i$ is inactive at $t$ ($y_i(t) = 0)$, active at $t'$ ($y_i(t') = 1)$, and the width of the interval $[k_i(t), k(t')] = 1$.
\label{condition1}
\end{condition}

In other words, if exposure changed by exactly one and this triggered $i$'s adoption, then we know the last neighbor to activate provided the ``final push'' for adoption.

Condition 1 has two edge cases. First, innovators who adopt with 0 active neighbors are measured precisely only by assumption. These nodes technically have no lower bound on their threshold intervals. Second, some nodes may have no lower bound, for instance a node who first updates and activates at exposure 2. Condition \ref{condition1} categorizes these nodes as imprecisely measured.

\begin{figure}[ht]
\includegraphics[width=\linewidth]{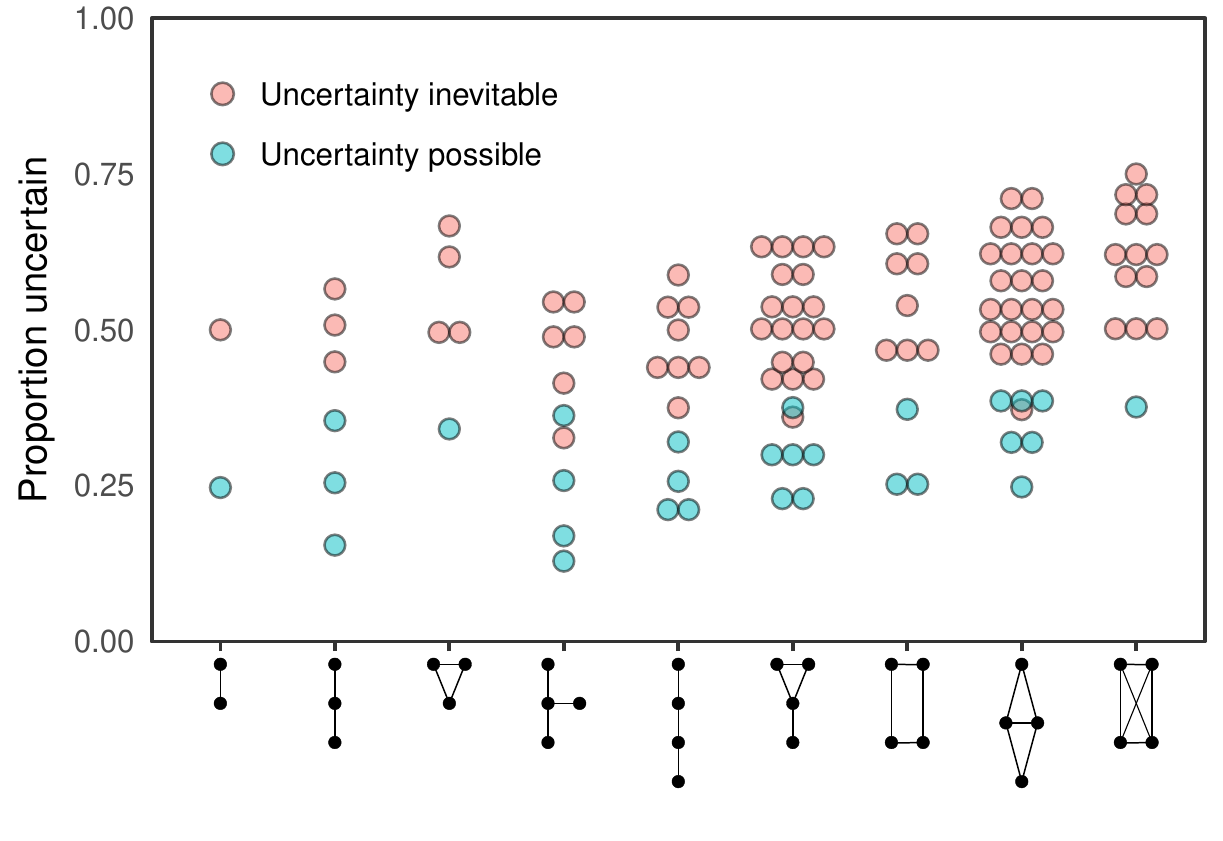}
\centering
\caption{The severity of the opacity problem in all connected graphs of order 2 to 4, with all threshold assignments that support diffusion. Each point represents a unique threshold assignment for the graph on the $x$-axis. The $y$-axis displays the mean number of nodes whose thresholds are uncertain when applying Condition \ref{condition1}. For 86/115 (75\%) of graph-threshold combinations, the EAA rule always produces at least one uncertain node (red dots). Since these small graph-threshold combinations are the beginnings of large cascades, we should expect measurement error when studying social contagion observationally.}
\label{figure22}
\end{figure}

In Figure \ref{figure22} we apply Condition \ref{condition1} to all small graphs of orders 2 to 4, with all threshold assignments that support diffusion. These graph-threshold combinations are the building blocks of large cascades. Figure \ref{figure22} indicates that for 75\% of these graph-threshold combinations, at least one node always has an interval which is not precisely measured according to Condition \ref{condition1}. This plot also shows the proportion of precisely measured nodes given nodes update randomly, indicating that measurement is difficult in many small graphs.

\section{Simulation evidence}

Simulating the threshold model in the previous section provides a method to examine the severity of the opacity problem under different assumptions \cite{macy_factors_2002}. In this case, thresholds are known with certainty which allows exact quantification of error. 

We use simulation to perform two analyses. First, the impact of opacity on observational threshold distributions is assessed. Then, we demonstrate that combining Condition \ref{condition1} with a predictive model such as linear regression can reduce the error in estimated thresholds.

Graphs are generated either with a power law with clustering (PLC) algorithm \cite{holme_growing_2002} with clustering parameter 0.1, or with a Barab\`{a}si-Albert (BA) procedure \cite{barabasi_emergence_1999}. All graphs have 1000 nodes, and have degrees ranging between 12 and 20. Each parameter set is replicated 1000 times to minimize sampling variance\footnote{We also examined Watts-Strogatz graphs but omit them for brevity since they did not meaningfully change the results.}.

The threshold model developed above is simulated in the following way. At each time step $t'$, choose an inactive $i$. Node $i$ checks its exposure $k_i(t')$, and we record this exposure regardless of whether $i$ activates. If $k_i(t') \ge h_i$, $y_i(t')$ is set to 1. If $i$ activates, its exposure at $t'$ is differenced from its exposure at $i$'s previous update at $t$, $k_i(t') - k_i(t)$. We apply Condition \ref{condition1}, and call $i$ ``precisely measured'' if $k_i(t') - k_i(t) = 1$. By assumption, nodes that activate with 0 active neighbors are precisely measured.

In addition to the threshold model, we also simulate the susceptible-infected (SI), or independent cascade model (ICM) \cite{hethcote_mathematics_2000,kempe_maximizing_2003,pastor-satorras_epidemic_2015}. We provide the simulation algorithms in the Appendix. Since the SI model is probabilistic, this analysis provides a robustness check against the deterministic nature of the threshold model. In the SI model, each active node provides an independent chance for its neighbors to adopt. We set this \emph{transmission probability} to $p = 0.2$. Two types of contagion mechanics are simulated for the SI model. First ``pull'' dynamics: when a node is selected to update, it checks how many newly active neighbors it has, flips that many coins with $P(\text{heads}) = p$, and activates if at least one comes up heads. Second, ``push'' dynamics: an active node is selected at random, and a single coin with P(\text{heads}) = p is flipped for each of its neighbors. Any neighbors that flip a heads are activated immediately. Each active node may only ``push'' the contagion once.

For the SI model, the true ``threshold'' or ``critical value'' for any node is unknown in advance. The simulation process reveals, for each node, a series of coin flips. We record the first such coin flip that comes up heads as the ``critical value'' or ``threshold'' for the node.

\subsection{Effect of opacity on threshold distributions}

The distribution of active neighbors recovered using the EAA rule systematically differs from the true threshold distribution. Figure \ref{fig444} shows the divergence using the Barab\`{a}si-Albert graph with mean degree 12. Distributional divergence is the case for integer thresholds, fractional thresholds, and the SI model. The case in Figure \ref{fig444}-D is surprising: all nodes have a fractional threshold of 0.2, yet the distribution recovered using the EAA rule is roughly uniform. In this case, the EAA rule records about 10\% of nodes as having ``threshold'' 1.0.

The EAA rule performs best in Figure \ref{fig444}-B, where the threshold distribution is normal. In this case, the EAA rule still produces a long tail of ``high threshold'' nodes, which gives the impression that a heroic level of peer reinforcement is required for some nodes to adopt.

These results indicate that opacity presents substantial challenges for estimating threshold distributions. Regression-based approaches relying on a biased distribution can have an influence coefficient biased in either direction. When the EAA rule moves observations from zero to nonzero thresholds, the strength of influence can be overestimated. On the other hand, when the EAA rule moves low threshold nodes to higher values, the effect of peer influence can be under-estimated. The particular bias is based on the relative strength of these two forces.

\subsection{Reducing measurement error with predictive models}

\begin{figure}[ht] 
\includegraphics[width=.8\linewidth]{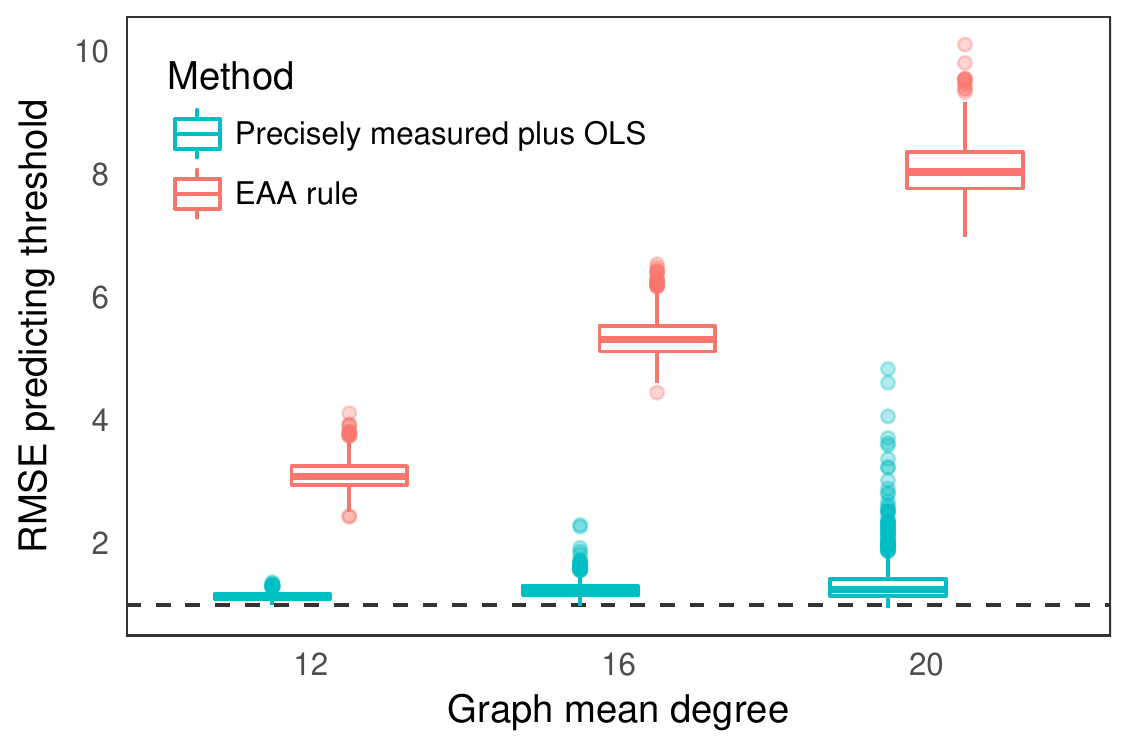}
\centering
\caption{Since the exposure-at-activation (EAA) rule takes the maximum of threshold intervals, it leads to over-estimates of thresholds (shown in red). The dashed black line gives the baseline RMSE, which is 1 in these simulations. The EAA rule leads to an average error between 3.1 and 8.1 times this baseline. A model estimated on the precisely measured subset ($< 15\%$ of nodes) and then used to predict all thresholds (blue) produces much lower error of 1.14-1.35 times baseline.}
\label{p5}
\end{figure}

\begin{figure}[ht]
\includegraphics[width=\linewidth]{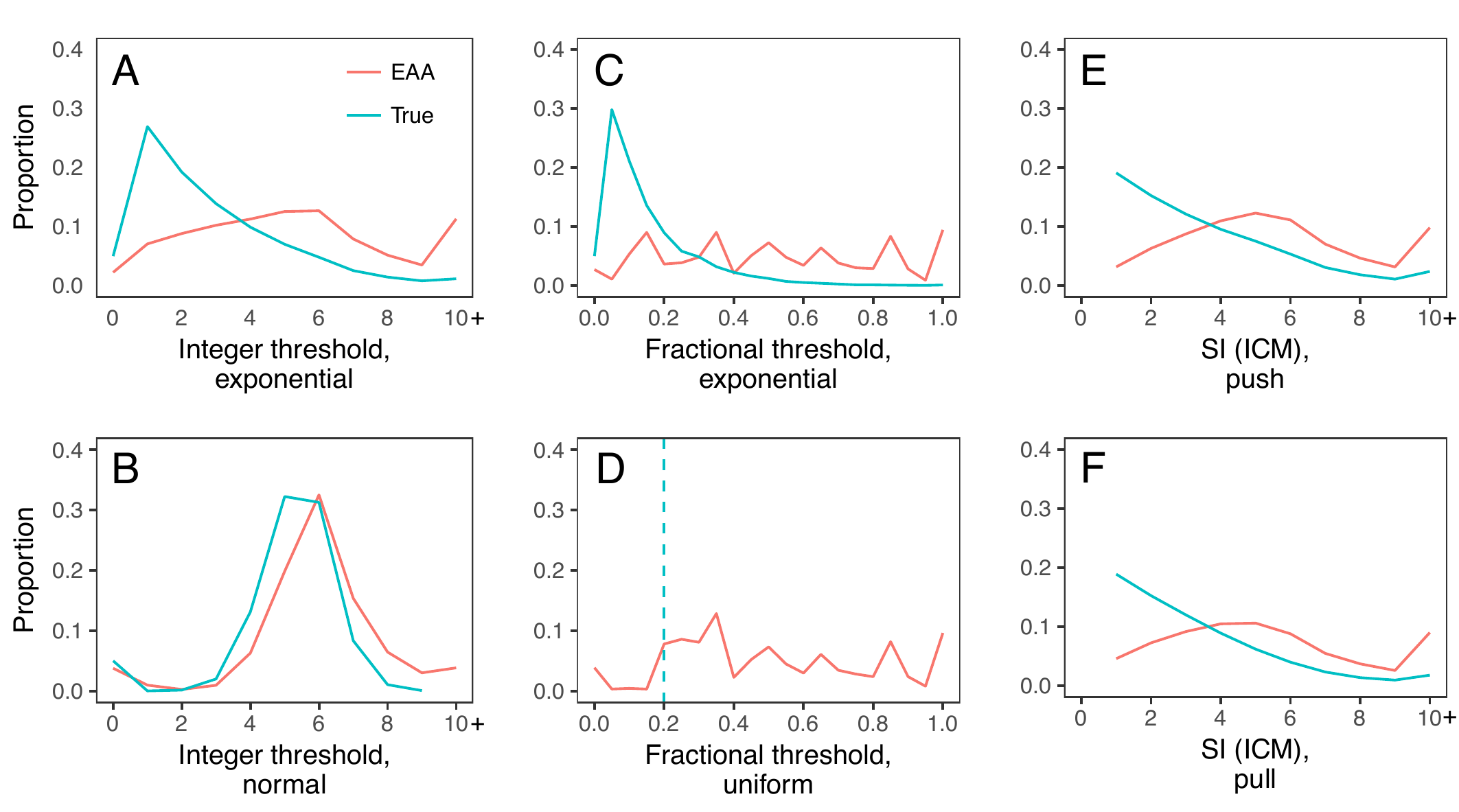}
\centering
\caption{True thresholds (blue) compared to measurements using the exposure-at-activation (EAA) rule (red) for a variety of contagion processes. A) and B) show the threshold model with integer thresholds drawn from an $\text{exp}(\beta=3)$ and $\mathcal{N}(5, 1)$, respectively. C) and D) use fractional thresholds (note the different $x$-axis scale) with 5\% seed nodes. C) uses an $\text{exp}(\beta=3)$ normalized to the $[0, 1]$ interval, while D) sets all non-seed nodes to have threshold 0.2. E) and F) use a susceptible-infected (SI) model, also called the independent cascade model (ICM), where each node has an independent chance $p = 0.2$ to activate peers. In this case, the ``threshold'' is the first coin flip to come up heads. In E), when a node activates it ``pushes'' contagion to its neighbors by causing them to activate with probability $p$ in a random order. In F), when nodes update they check all neighbor statuses and flip a probability $p$ coin for each newly activated neighbor.}
\label{fig444}
\end{figure}

Bias in thresholds can be meaningfully reduced using predictive models. The strategy is straightforward: apply Condition \ref{condition1} to all nodes, determine which nodes are measured precisely, and estimate a model on these nodes only. We use Ordinary Least Squares (OLS) regression, although in principle any machine learning regressor can be used. Then, we predict thresholds for nodes with measurement error. As shown in Figure \ref{p5} Results indicate this approach reduces root mean squared error (RMSE) dramatically, from 3.1 to 8.1 times baseline to 1.15 to 1.35 times baseline.

We use a power law with clustering graph and threshold function $h_i = 5 + 3x_i + \epsilon_i$, where $x_i, \epsilon_i \sim \mathcal{N}(0, 1)$. Thresholds below 0 are set to 0. Cascade dynamics are simulated as described above. The covariate $x_i$ represents some feature of $i$ (e.g. age, wealth) that provides information about how much social reinforcement $i$ needs before activation.

Treating thresholds as a function of node-level attributes \cite{valente_social_1996} allows estimating a threshold model

\begin{equation}
h_i = \alpha + \beta x_i + \epsilon_i
\end{equation}

This model can be estimated using only the precisely measured set of nodes to reduce bias in thresholds. Results of this procedure compared to the EAA rule are displayed in Figure \ref{p5}. We note that this model predicts thresholds based on node attributes, rather than on endogenous network characteristics.

The error variance represents the inherent unpredictability in thresholds and provides a natural baseline error for comparing error-reduction methods. An ordinary least squares (OLS) model estimated on the true data will have a root-mean-square error (RMSE) equal to the error variance, which here is set to 1.

\begin{table}
\centering
\begin{tabular}{cccc}
  \hline
  \hline
    Mean degree & Nodes & Activations & Precisely measured \\ 
  \hline
    12 & 1000 & 720.7 & 113.4 \\
    16 & 1000 & 924.1 & 43.0  \\
    20 & 1000 & 983.1 & 13.0 \\
  \hline
\end{tabular}
\caption{Descriptive statistics on simulations, averaged over 1000 runs. The average number of activated nodes increases as degree increases. However, it becomes more difficult to precisely measure nodes in higher mean degree graphs.}
\label{table1}
\end{table}

\section{Hashtag cascades}

Studying hashtag cascades on Twitter, we find that the opacity problem can qualitatively change conclusions about the effects of peer reinforcement on adoption. Twitter provides an ideal setting for this analysis: tweets (and therefore adoption events) are timestamped and the contagion (a hashtag) can be clearly tracked. For about 2 in 5 hashtags, different responses to opacity yield opposite conclusions about the effectiveness of social reinforcement. At the adoption level, about 1 in 5 hashtag adoptions show uncertainty about the amount of peer reinforcement required for adoption.

We tracked cascades for 50 hashtags (see Appendix for list and selection procedure) with between 40,000 and 360,000 unique adopters and 41,000 to 1.3 million total usages per tag. Retweets were filtered out. A total of 3.2 million users (``active users'') tweeted using one or more of these hashtags. These active users are connected by 45 million bidirected @mention links, which is a common measure of ties on Twitter \cite{romero_differences_2011}. An additional 105 million bidirected @mention edges connect active users to users who were exposed but never adopted (``inactive users''). These data contain a total of 7 billion tweets.

We choose to construct the network from bidirected @mentions because they represent a relatively strong connection that can facilitate diffusion. This type of link is sparser than other types of edges on Twitter, such as follows. Combined with our findings above about the positive correlation between network density and severity of the opacity problem, using the sparser bidirected @mention graph can be considered a conservative test of the impact of opacity.

We examine the effect of peer reinforcement with $p(k)$ curves \cite{crandall_feedback_2008,romero_differences_2011,lerman_information_2016}, which plot the probability of first activation with exposure $k$, given ever being $k$-exposed. Work using this methodology has found that higher exposure levels substantially increase the probability of hashtag adoption, particularly for controversial issues like politics \cite{crandall_feedback_2008,romero_differences_2011}. More recent work has found that hashtags spread more like simple contagion where nodes are immunized after the first exposure \cite{lerman_information_2016}. Since complex contagions have different cascade dynamics than simple contagions \cite{centola_cascade_2007,barash_critical_2012}, understanding the effects of reinforcement is consequential for predictions and interventions in online networks.

\subsection{Constructing threshold intervals}

To construct threshold intervals, we assume updates correspond with tweet times. While this is not a perfect assumption, a user is engaged with the Twitter platform when tweeting, and likely checks tweets in the seconds before or after sending a tweet. Let $\mathbf{u}_i$ be the times $i$ tweeted in increasing order. For hashtag $g$, let $t_i^g \in \mathbf{u}_i$ be $i$'s first usage of $g$, and $t_i^{g-1} \in \mathbf{u}_i$ be the tweet immediately prior to $i$'s first usage of $g$.

We construct descending time intervals $(t_i^{g- 1}, t_i^g], (t_i^{g - 2}, t_i^{g - 1}]$, etc., which give the periods of time between $i$'s updates. Iterating through these intervals, we find the most recent interval for which one or more neighbor adopted $g$, and count the total number $T^*$ of neighbor adoptions in that interval. If $T^* = 1$, then Condition \ref{condition1} is satisfied, since $i$'s exposure changed by exactly one and then $i$ adopted. Otherwise if $T^* > 1$, $i$'s threshold is uncertain within an interval of size greater than 1. For instance if 2 neighbors activated in $(t_i^{g-1}, t_i^g]$ and 4 total neighbors activated before $t_i^{g-1}$, then $i$'s threshold interval is $[4, 6]$. This interval is considered ``imprecisely measured'' since it has size greater than 1.

\subsection{Empirical measurement rates}

Measurement uncertainty affects 21\% of activations, although there is substantial variation both by hashtag and exposure level.  We exclude innovators (0 active neighbors at activation) from this analysis both to focus on the effects of opacity on estimates of peer reinforcement. Nodes which adopt with higher exposure are more likely to suffer from measurement uncertainty \ref{table4}. Among nodes that activate with exposure 1, 91\% are measured precisely. Among nodes that activate with exposure 2, only 66\% are precisely measured. For each hashtag, we compute the clustering coefficient \cite{wasserman_social_1994} among the induced subgraph of adopters, and display measurement rates for the top and bottom quartile of hashtags.

\begin{table}[ht]
\centering
\begin{tabular}{r|ccc}
  \hline
  \hline
  & All & Low clustering & High clustering \\ 
\hline
      Exposure & \multicolumn{3}{c}{Proportion precisely measured} \\
      \hline
  1 & 0.91 & 0.90 & 0.97 \\ 
  2 & 0.66 & 0.74 & 0.46 \\ 
  3 & 0.64 & 0.70 & 0.43 \\ 
  4 & 0.63 & 0.68 & 0.41 \\ 
  5 & 0.63 & 0.66 & 0.42 \\ 
\hline
  N hashtags & 50 & 13 & 13 \\
  N adoptions & 1,625,759  & 548,420 & 262,890 \\
  \end{tabular}
\caption{Descriptive statistics for Twitter hashtag cascades. There are 1.6 million adoption events where exposure is between 1 and 5 at the time of adoption. Precise measurement rates are higher for adoptions with exposure 1 than higher exposure levels. Splitting hashtags into quartiles of clustering coefficient among adopters, low and high clustering tags display substantially different precise measurement rates.}
\label{table4}
\end{table}

\subsection{Social reinforcement and $p(k)$ curves}

We construct two $p(k)$ curves for each hashtag based on the threshold intervals for each node. The ``upper'' $p(k)$ curve, notated $p_U(k)$, takes the maximum of each threshold interval for each activation. This is equivalent to applying the EAA rule found in much work on social contagion. The ``lower'' $p(k)$ curve, notated $p_L(k)$, takes the minimum of each threshold interval for each activation. In other words, it assumes that the EAA rule is maximally wrong. While taking the minimum of each threshold interval may seem like an unrealistic assumption, it has two advantages: 1) the lower curve does not have the long right tail present when using the EAA rule (as seen in Figure \ref{fig444}); 2) assuming the EAA rule is maximally wrong allows us to bound the true value in an interval, assuming that peer influence was the actual reason for adoption.

This analysis is presented in Figure \ref{mp3}. We find that nearly all hashtags had initially increasing upper curves, $p_U(2) > p_U(1)$, corresponding to social reinforcement facilitating adoption and qualitatively agreeing with past work \cite{crandall_feedback_2008, romero_differences_2011}. When turning to lower curves, we noticed via visual inspection that many hashtags had the opposite pattern, $p_L(2) < p_L(1)$, where the probability of adoption drops after the first exposure. This pattern indicates that contagion spreads more like an independent cascade, with exposure leading to immunization (as proposed in \cite{lerman_information_2016}). We find that for 19 out of 50 hashtags (full list in Appendix), the upper curve displays the usual pattern of increasing adoption probability in $k$, while the lower curve displays the decreasing pattern. For the remaining 31 hashtags, both upper and lower curves indicate adoption probability increasing in $k$, with a smaller increase for the lower curve.

We examine three characteristics of cascades which could be correlated with discrepancies between the upper and lower curves: 1) the total number of hashtag adopters; 2) the clustering coefficient among the induced subgraph of adopters; 3) the temporal burstiness of the cascade, measured by the Gini coefficient of the number of adoptions across days between the first and last usages of a particular hashtag. Of these three factors, clustering coefficient and burstiness have correlations with a decreasing lower curve (Pearson correlation for clustering coefficient: 0.26, $p$ = 0.063, for Gini coefficient: 0.34, $p$ = 0.017). Hashtags with high clustering coefficients tend to have particularly pronounced differences between upper and lower curves, as seen in Figure \ref{mp3}-B. 

Taken together, these results indicate that the opacity problem creates more uncertainty about cascade dynamics when adopters are clustered in dense networks and cascades happen in short time spans. This paradoxically makes it difficult to determine the effects of social reinforcement when adopters have high clustering, even though contagion which requires social reinforcement is most likely to spread in clustered networks \cite{centola_complex_2007}.

\section{Implications and future work}

The opacity problem and the application of the EAA rule present several challenges for the study of contagion and social influence. These can be summarized into the following categories: 1) estimating individual thresholds and the distribution of thresholds; 2) assessing impacts on empirical models of contagion; 3) using data to better assess uncertainty.

The first challenge we have substantially addressed in this paper. Future work can build on this by assessing the information contained in low-error observations. We have only considered observations measured precisely, but an observation with a threshold interval of size 2 is much more certain than one with size 20. Bayesian and machine learning techniques may prove useful in this direction. For instance, formulating the labeling problem as a semi-supervised task could lead to lower error threshold predictions than the method we have proposed. In addition, the predictive value of various signals remains to be assessed empirically. For instance, in the Twitter case, using text embeddings \cite{mikolov_distributed_2013} representing cultural taste and graph embeddings \cite{grover_node2vec:_2016} indicating attention may prove useful for predicting thresholds.

The second category poses a series of issues that we have not touched upon in this paper: the specific implications of using exposure as an independent variable in a model estimating peer effects on adoption. A simple version of this model can be written $y_i(t) = \beta k_i(t) + \epsilon_i$. In this case, the opacity problem plus EAA rule presents a difficult errors-in-variables problem: $k_i(t)$ is measured with error \textit{conditional on} $y_i(t) = 1$, with non-negative error for all nodes $i$. Depending on the specific form measurement error takes, $\hat{\beta}$ may be over- or under-estimated. Future work can assess the impacts of opacity in a variety of empirical cases.

The third challenge takes two forms. When data about adoption times is relatively granular (as in the case of social media), methodology proposed here can be applied to assess the impacts of opacity. This is straightforward, and doing so will provide insight both about the substantive social process and the difficulty of measuring various diffusion processes. When data is not granular, as in the case of surveys which may be administered yearly, care needs to be taken to assess opacity. To address the problem, a yearly survey may consider asking about the adoption of a behavior and about the specific time that behavior was adopted, allowing a partial reconstruction of the adoption ordering among network neighbors. This could reduce the severity of the opacity problem and lead to better inferences about social influence.

We are aware of one case where the opacity problem can be avoided entirely: ego-network randomization\cite{bakshy_social_2012,aral_identifying_2012}. In this experimental design, the researcher can control the number of active neighbors displayed, leading to valid estimates of adoption probabilities at various exposure levels.

Recognizing and addressing the opacity problem has the potential to yield more precise answers to important questions. For instance, models of node critical values can be used to better study complex contagion \cite{centola_complex_2007} empirically. Results from such an analysis can then be applied to problems such as influence maximization \cite{kempe_maximizing_2003,kempe_influential_2005} and the cascade prediction problem \cite{cheng_can_2014}.

\section{Figures}



\begin{figure}[ht]
\includegraphics[width=\linewidth]{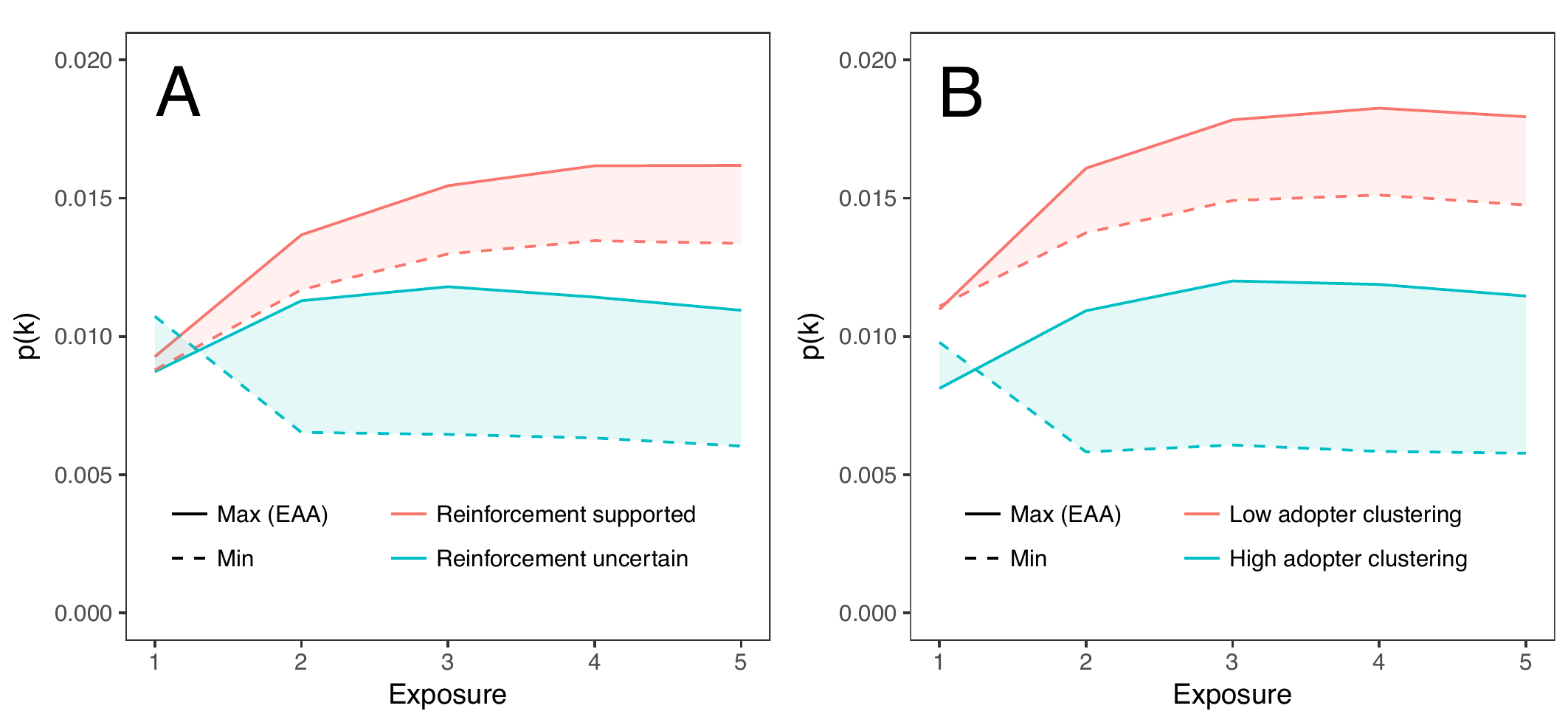}
\centering
\caption{Different responses to the opacity problem support different conclusions of the importance of peer reinforcement for adoption. This can be seen with $p(k)$ curves, which plot the probability of first activation at $k$-exposure given ever being $k$-exposed. Solid lines indicate using the exposure-at-activation (EAA) rule (the maximum of threshold intervals). Dashed lines indicate using the minimum of threshold intervals. The shaded region is the discrepancy between the two methods. A) shows that for 19/50 hashtags (blue), the upper and lower curves lead to different conclusions about the importance of reinforcement: use of the EAA rule supports a complex contagion hypothesis, while taking the minimum of threshold intervals supports an independent contagion hypothesis. B) shows that this pattern can be replicated by taking hashtags with high and low clustering in the bidirected @mention graph among adopters, indicating that the dynamics of hashtags in highly clustered networks are particularly uncertain.}
\label{mp3}
\end{figure}

\FloatBarrier

\section{Acknowledgements}

We thank David Strang, Thomas Davidson, Carter Butts, Tom Valente, participants of INSNA Sunbelt 2016, participants of the 2017 Annual Meeting of the American Sociological Association, and members of the Social Dynamics Lab for helpful discussions and comments. This research has been supported in part by National Science Foundation grant SES-1357442 and the Department of Defense Minerva Initiative grant FA9550-15-1-0036.

\section{Appendix}

\subsection{Simulation details}

We simulate cascades using the NetworkX package in Python 3 \cite{hagberg_exploring_2008}. All code is available at \url{https://github.com/georgeberry/thresholds}. Statistical analyses are done using Scikit Learn \cite{pedregosa_scikit-learn:_2011} in Python 3, and in R.

We use 4 different graph generation routines at various points: Barab\`{a}si-Albert \cite{barabasi_emergence_1999}, power-law with clustering \cite{holme_growing_2002}, Watts-Strogatz \cite{watts_collective_1998}, and an ``atlas'' of all small graphs \cite{read_atlas_1998}. All four are built-in to the NetworkX package.

\subsubsection{Threshold model}

The algorithm for simulating diffusion is as follows.

At each time step $t'$,

\begin{enumerate}
\item An inactive node $i$ is chosen at random
\item Node $i$ checks how many active neighbors it has, $k_i(t')$
\item The exposure at $t'$ is recorded
\item If $k_i(t') \ge h_i$, set $y_i(t') = 1$
\item If $i$ activates at $t'$
\begin{enumerate}
\item Apply Condition \ref{condition1} by differencing the exposure between $t'$ and $i$'s previous update at $t$, $k_i(t') - k_i(t)$
\item If $i$ did not update at any previous $t$, we set node $i$ to ``imprecisely measured'' unless node $i$ activated with exposure 0, in which case we set it to ``precisely measured''
\item If $k_i(t') - k_i(t) = 1$, set node $i$ to ``precisely measured'', otherwise ``imprecisely measured''
\end{enumerate}
\end{enumerate}

End iteration when A) all nodes have activated; B) each inactive node has been checked consecutively without activating.

\subsubsection{SI/ICM ``pull'' model}

This model is similar to the threshold model, except activations are decided by coin flips instead of deterministically.

At each time step $t'$,

\begin{enumerate}
\item An inactive node $i$ is chosen at random
\item Node $i$ checks how many active neighbors it has, $k_i(t')$
\item $i$ checks the difference between its exposure at last update $t$ and present, $d_i= k_i(t') - k_i(t)$
\item $i$ flips $d_i$ coins, each of which has probability $P(\text{heads}) = 0.2$
\item If the $j$th coin comes up heads, activate $i$ and set $i$'s ``threshold'' to $k_i(t) + j$ (previous exposure plus coins flipped at $t'$ before getting heads)
\item If $i$ activates at $t'$
\begin{enumerate}
\item Apply Condition \ref{condition1} by differencing the exposure between $t'$ and $i$'s previous update at $t$, $k_i(t') - k_i(t)$
\item If $i$ did not update at any previous $t$, we set node $i$ to ``imprecisely measured'' unless node $i$ activated with exposure 0, in which case we set it to ``precisely measured''
\item If $k_i(t') - k_i(t) = 1$, set node $i$ to ``precisely measured'', otherwise ``imprecisely measured''
\end{enumerate}
\end{enumerate}

End iteration when A) all nodes have activated; B) each inactive node has been checked consecutively without activating.

\subsubsection{SI/ICM ``push'' model}

In this model active nodes are selected to update, and they get a single chance to activate each one of their neighbors.

Initialize an empty array of active nodes that have updated, $a$.

At each time step $t'$,

\begin{enumerate}
\item An active node $i$ not in $a$ is chosen at random and appended to $a$
\item The neighbors of $i$ are randomized, $\text{shuffle}(N(i))$ 
\item For each $j$ in $\text{shuffle}(N(i))$ 
\begin{enumerate}
\item Node $j$'s exposure $k_j(t')$ is recorded
\item Node $j$ flips a coin with $P(\text{heads}) = 0.2$ and activates if it comes up heads
\item If $j$ activates at $t'$
\begin{enumerate}
\item Apply Condition \ref{condition1} by differencing the exposure between $t'$ and $j$'s previous update at $t$, $k_j(t') - k_j(t)$
\item If $i$ did not update at any previous $t$, we set node $j$ to ``imprecisely measured'' unless node $j$ activated with exposure 0, in which case we set it to ``precisely measured''
\item If  $k_j(t′)−k_j(t)=1$ , set node $j$  to ``precisely measured'', otherwise ``imprecisely measured''
\end{enumerate}
\end{enumerate}
\end{enumerate}

End iteration when A) all nodes have activated; B) all active nodes are in $a$.

\subsubsection{Small graphs}

NetworkX provides a full enumeration of all small graphs (``graph atlas''). We take all such graphs with between two and four nodes. Then, we filter out graphs which have more than one component, giving a set of all connected graphs with between two and four vertices. Call this set of graphs $G$.

For each $g \in G$, we generate all possible critical value assignments given that the critical value is less than or equal to node degree. For node $i$ in graph $G$ with degree $d_i$, $i$ has the critical value set $H_i = \{h_i : h_i \le d_i\}$. Taking the product of $H_i$ for all $i \in g$ gives the set of critical value assignments for the graph $H(g)$.

For each critical value assignment, we simulate cascades by updating nodes randomly and activating nodes immediately if exposure is greater than critical value, $k_i \ge h_i$. We filter out graphs where at least one node never activates, which occurs when each inactive node has updated yet none activates.

For these set of ``admitted'' graphs $G^*$ where all nodes eventually activate, we simulate 100 cascades per graph. For each simulated cascade, we record exposure at each node update. This allows applying the exposure-at-activation (EAA) rule to determine if nodes are precisely measured or not. If node $i$ updates at $t$ and some later $t'$, and $k_i^t + 1 = k_i^{t'}$, then the node is precisely measured. If there is no lower bound (e.g. no update before activation) and the exposure at activation is greater than zero, we call the node imprecisely measured. Innovators which adopt with 0 active neighbors are considered precisely measured here. As we discuss in the main text, this assumption about innovators may not be appropriate in all cases.

\subsection{Twitter analysis}

The Twitter data used for this analysis was collected for another project via the Twitter REST API between November 2013 and October 2014. This was supported by an NSF grant (SES 1226483). Tweets were localized to a country using the method described in \cite{compton_geotagging_2014}. Users in Anglophone countries (US, UK, CA, AU, NZ, SG) were extracted and retweets were filtered out. We selected users who had used one or more of 50 hashtags with moderate-to-high usage. We analyzed the entire timeline of these users which was collected during the data collection process.

\subsubsection{Hashtag selection}

Hashtags were selected in the following way. The first occurrence of the hashtag in the data was between 2012 and 2014. Two authors (G.B. and C.C) manually examined the top 1000 hashtags by usage, and independently nominated tags which they believed were related to 1) specific offline events, such as \#riprobinwilliams; 2) social media phenomena, such as \#windows8. The nominated tags were pooled and the authors discussed disagreements.

The list of lowercased tags which are classified as ``reinforcement supported'' in Figure \ref{mp3} is: benghazi, bringbackourgirls, cantbreathe, drawsomething, election2012, euro2012, firstvine, gop2012, harlemshake, ios6, jodiarias, justicefortrayvon, linsanity, marriageequality, miley, nfldraft2014, nobama, obama2012, romney, romney2012, romneyryan2012, samelove, snowden, springbreak2014, teamobama, trayvon, trayvonmartin, voteobama, whatdoesthefoxsay, windows8, zimmerman

The list of lowercased tags which are classified as ``reinforcement uncertain'' in Figure \ref{mp3} is: betawards2014, debate2012, ferguson, goodbyebreakingbad, governmentshutdown, hurricanesandy, inaug2013, ivoted, kony2012, mentionsomebodyyourethankfulfor, newtown, olympics2014, prayersforboston, prayfornewtown, replaceashowtitlewithtwerk, rippaulwalker, riprobinwilliams, sharknado2, worldcupfinal

\bibliography{library}
\bibliographystyle{plain}

\end{document}